\documentstyle[psfig,conf_iap,]{article}
\newcommand{\be}{\begin{equation}}
\newcommand{\ee}{\end{equation}}
\def\vh{\varphi}
\begin{document}
\heading{%
%
Relative Standard of Measurement and Supernova Data
%
}
\par\medskip\noindent
\author{%
David Blaschke$^{1,2}$, Danilo Behnke$^{1}$, Victor Pervushin$^2$, 
Denis Proskurin$^2$
}
\address{%
  University of Rostock, Department of Physics,
  Universit\"atsplatz 3, 18051 Rostock, Germany.
}
\address{%
  Bogolyubov Laboratory for Theoretical Physics, Joint Institute for Nuclear
Research,
  141980 Dubna, Russia
}
\begin{abstract}
We show that in the units of the relative Paris meter both the
latest data on the Supernova luminosity-distance -- redshift relation
and primordial nucleosynthesis are described by the dynamics of
a homogeneous, massless scalar field (Scalar Quintessence) in a nonexpanding
universe.
\end{abstract}

\section{Introduction}
The analysis of the magnitude -- redshift relation from observation of distant
type Ia Supernovae \cite{snov,1997ff} shows that these standard candles have 
larger luminosity distances than expected for a flat, matter-dominated 
universe.
In terms of standard cosmology (SC), which explains the redshift of spectral 
lines by the Doppler effect of receding galaxies, a perfect fit of the data is 
obtained for a nonvanishing positive cosmological constant 
$\Omega_{\Lambda} = 0.7$ (dark energy) of the same order of magnitude as the 
matter density $\Omega_{\rm M} = 0.3$, i.e. for an accelerated expansion of 
the universe.
The origin and the magnitude of the dark energy is the puzzle which brings us 
together at this conference. 
As was discussed e.g. by N. Straumann \cite{stra}, the fact that condensates 
in quantum field theories lead to a $\Lambda$ term orders of magnitude larger 
than the observed one points to a deep problem.
Note, however, that the cosmological constant problems are intimately related 
with its dependence on the volume of the expanding universe. 
Most of the attempted solutions suggest a time-dependence of fundamental 
quantities, like quintessence \cite{quint}, varying alpha \cite{alpha} or 
varying speed of light theories \cite{light}.

In the present contribution, we want to discuss the alternative of a conformal 
cosmology (CC) in which the universe is not expanding and the observed redshift
is caused by the time-dependence of all particle masses \cite{039}.

\section{Relative Standard of Measurements}

Astrophysical data are described in  Einstein's general relativity (GR)
\begin{equation}\label{gr1}
  S_{\rm GR}[\vh_0|g]=-\int d^4x\sqrt{-g}~\frac{\vh_0^2}{6}~R(g)
\end{equation}
using as an absolute standard of measurement the Newtonian gravitational 
constant $G_0=6/{\vh_0^2}$ and the interval (flat space) 
$(ds^2)=g_{\mu\nu}dx^{\mu}dx^{\nu}=(dt)^2-a^2(t)(dx^i)^2$.
This metrics describes an expanding  spatial volume of the universe
$V_{\rm SC}(t)=V_{\rm CC} a^3(t)$ and all lengths in
the universe are measured with respect to the absolute Paris meter
\be\label{apm}
{\rm Absolute~Paris~Meter} =1 {\rm m}.
\ee
However, this absolute Paris meter cannot be realised since it does not exist 
outside the universe and is subject to the cosmological changes of length 
scales.
Instead, we have to use a relative Paris meter
\be\label{rpm}
{\rm Relative~Paris~Meter} =1 {\rm m}\times a(t)
\ee
for the measurements of {\it all lengths} with the corresponding conformal 
line element
$ds^2/a^2(t)=(d\eta)^2-(dx^i)^2$ which is Minkowskian and not expanding.
The measurable spatial volume of the universe $V_{\rm CC}$ is a
constant while the measurable Planck mass
\be\label{abs}
M_{\rm Planck}(\eta)=\vh(\eta) \sqrt{8\pi \hbar c/3}
=2.177\times 10^{-8}{\rm kg}\times a(t),
\ee
in the relative units becomes a dynamic variable of the conformal time
$d\eta=dt/a(t)$,
\be\label{1pct2}
\vh(\eta)=\vh_0 \times a(t),
\ee
as all measurable masses of elementary particles 
$m_{\rm CC}(\eta)=m_{\rm SC} \times a(t)$ .
The spectrum of photons emitted by atoms from distant stars billion
years ago remains unchanged during the propagation and is determined by the 
mass of the constituents at the moment of emission. 
When this spectrum is compared with the spectrum of similar atoms on the Earth
which, at the present time, have larger masses then a redshift is obtained.

The conformal density is $\rho_{\rm CC}=\rho_{\rm SC}\times a^4(t)$, and the 
conformal temperature $T_{\rm CC}=T_{\rm SC} a(t)={\rm const}$.
The common point for the two cosmologies is the identification of the
evolution of the universe with the  evolution of the cosmic scale factor
$a(t)= (1+z)^{-1}$ which in both cosmologies has a different meaning
\cite{039,ppgc,bier}.
In the standard cosmology the cosmic factor scales all distances {\it besides
the Paris meter}~(\ref{apm}). 
In the conformal cosmology it scales  all masses {\it including the Planck 
mass}~(\ref{abs}).
As it was shown in~\cite{039}, in the case of the {\it relative} Paris meter~
(\ref{rpm}), both the recent experimental data for distant supernovae 
\cite{snov} are described by the evolution according to a rigid equation of 
state, see Fig.1,
\be\label{data3}
 (z+1)^{-1}(\eta)=\sqrt{1+2H_0(\eta-\eta_0)}.
\ee
This evolution results from the dynamics of a homogeneous scalar field with 
the action~\cite{ppgc}
\be\label{SQ1}
S_{\rm SQ}[\vh_0|{g}]=\int d^4x\sqrt{-g}~\vh_0^2~\partial_\mu Q\partial^\mu Q~,
\ee
which we call scalar quintessence (SQ). 
This massless field with purely kinetic contribution to the energy density in 
the universe leads to a rigid equation of state~(\ref{SQ1}) and gives a 
satisfactory description of the supernova data.
\begin{figure}[htb]
\centerline{\vbox{
\psfig{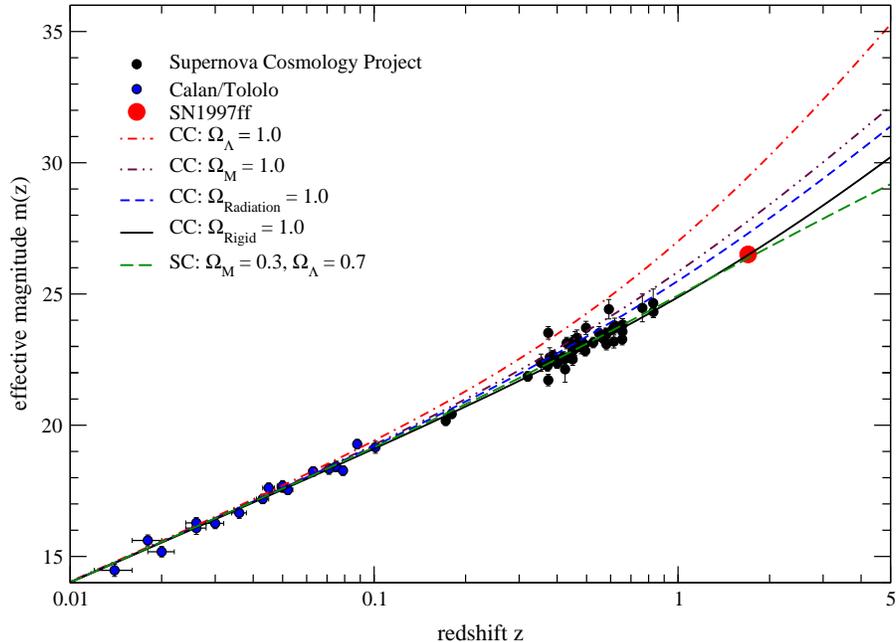}
}}
\caption[]{Luminosity-redshift-relation for a flat universe model in
  SC an CC. The data points include those from 42 high-redshift
  supernovae type Ia \cite{snov} and that of the recently reported
  farthest supernova SN1997ff \cite{1997ff}. An optimal fit to these
  data within the SC requires a cosmological constant
  $\Omega_{\Lambda}=0.7$, whereas in the CC these data require the
  dominance of the rigid state.}
\end{figure}

\section{Discussion}
Many arguments have been tried to disproof the concept of a nonexpanding
universe, we want to discuss some of them in concluding our contribution.
One of the consequences of the relative standard of measurement is the redshift
independence of the cosmic microwave background temperature \cite{pb}.
This is at the first glance in striking contradiction with the observation
\cite{sria} of $6.0 ~{\rm K} < T_{\rm CMBR}(z=2,3371) < 14~{\rm K}$.
The relative population of different energy levels $E_i$ from which the
temperature has been inferred in this experiment follows basically a
Boltzmann statistics.
The argument of these Boltzmann factors, however, has in the cooling
universe scenario and in the raising mass universe scenario the same
$z$-dependence \cite{039}. Therefore, the experimental finding can equally well
be interpreted as a measurement of the $z$-dependence of energy levels (masses)
at constant temperature.
A second question concerns bounds from nucleosynthesis. 
Although we have not performed a detailed calculation, we may answer this 
question in a similar way as the previous one. 
The abundances of nuclear species is also mainly governed by Boltzmann factors 
with an argument which is invariant with respect to the change of the picture: 
$m(z)/T(0)=m(0)/[(1+z)T(0)]=m(0)/T(z)$.
Since the time dependence of $z$ in the nonexpanding universe scenario follows
the radiation stage behavior, we expect the element abundances to be reproduced
well.

A last question we want to address concerns the fact that the supernovae
lightcurves are time-dilatated with exactly the same redshift that
is inferred from their spectra \cite{dilatation}.
In the expanding universe scenario this effect is a result of the stretching of
all length scales in the same way as Doppler effect causes the redshift.
In our approach, redshift originates from a scaling of masses, all masses
of elementary particles change with time. This entails changes in the
wavelengths of atomic spectra (therefore the redshift) as well as changes
in the ranges of interactions and the decay times for e.g. the beta decay
of Nickel which determines the decline of the type Ia supernovae light curves
and thus their duration~\cite{riaz}.
When we take into account that the decay time is proportional to the Fermi 
constant which is inversely proportional to the W-boson mass, the actual value 
of which is determined by the redshift, then we have the wanted relation 
between duration of supernova and redshift.

As a next step, we will consider the constraints from the fluctuations of the 
CMBR when the recombination era is considered within the variable mass scenario
of the nonexpanding universe discussud in the present contribution.

\end{document}